\def\year{2021}\relax
\documentclass[letterpaper]{article} % DO NOT CHANGE THIS
\usepackage{aaai21}  % DO NOT CHANGE THIS
\usepackage{times}  % DO NOT CHANGE THIS
\usepackage{helvet} % DO NOT CHANGE THIS
\usepackage{courier}  % DO NOT CHANGE THIS
\usepackage[hyphens]{url}  % DO NOT CHANGE THIS
\usepackage{graphicx} % DO NOT CHANGE THIS
\usepackage{natbib}  % DO NOT CHANGE THIS AND DO NOT ADD ANY OPTIONS TO IT
\usepackage{caption} % DO NOT CHANGE THIS AND DO NOT ADD ANY OPTIONS TO IT
\title{Human in the Loop for Machine Creativity}
\author{
    Neo Christopher Chung\textsuperscript{\rm 1}\thanks{This work was in part supported by CHIST-ERA [CHIST-ERA-19-XAI-007] to INFORM, by Narodowe Centrum Nauki [2020/02/Y/ST6/00071], and by OP ENHEIM Digital Residency.}
    \\
}
\affiliations{
    \textsuperscript{\rm 1}Institute of Informatics, University of Warsaw \\
       Stefana Banacha 2 \\ 
       02-097 Warsaw, Poland  \\
       nchchung@gmail.com}
\begin{document}

\maketitle

\begin{abstract}
Artificial intelligence (AI) is increasingly utilized in synthesizing visuals, texts, and audio. These AI-based works, often derived from neural networks, are entering the mainstream market, as digital paintings, songs, books, and others. We conceptualize both existing and future human-in-the-loop (HITL) approaches for creative applications and to develop more expressive, nuanced, and multimodal models. Particularly, how can our expertise as curators and collaborators be encoded in AI models in an interactive manner? We examine and speculate on long term implications for models, interfaces, and machine creativity. Our selection, creation, and interpretation of AI art inherently contain our emotional responses, cultures, and contexts. Therefore, the proposed HITL may help algorithms to learn creative processes that are much harder to codify or quantify. We envision multimodal HITL processes, where texts, visuals, sounds, and other information are coupled together, with automated analysis of humans and environments. Overall, these HITL approaches will increase interaction between human and AI, and thus help the future AI systems to better understand our own creative and emotional processes.
\end{abstract}

\section{Introduction}
\label{introduction}
As artificial intelligence (AI) becomes increasingly capable of recognizing and generating patterns and styles, its creative applications have led to creative visuals, texts, audio, and other data. Recently, deep neural networks have been used to produce artworks, free from or with minimal human interventions \cite{Gatys2016, Elgammal2017, GANEdmondBelamy}. However, the potential of human-computer interaction for creative application of AI has not been seriously considered. Here, we explore human in the loop (HITL) approaches for machine creativity, propose novel interactions, and discuss their implications. Beyond endowing algorithms to take more agency and creativity (even in limited definitions), this process will also help train the models for multi-modal data and nunanced relationships.

Digital prints by generative adversarial networks (GANs) \cite{Goodfellow2014} and variations are being auctioned \cite{GANEdmondBelamy} at premier art institutions. Texts generated by long short-term memory (LSTM) \cite{Hochreiter1997} and GPT-3 \cite{Brown2020} are being published as novels \cite{Goodwin2018} and textbooks \cite{BetaWriter2019}, respectively. Yet, outputs from state of the art models are often distinct, giving away their signature characteristics. GAN images often include blurring, shearing, and aliasing. More recent variations often result in unstructured non-hierarchical montages (e.g., dreamy images). AI-generated texts can be coherent within a limited scope. However, it’s much harder, if not impossible, to expect long-term structure, genuine portrayal of emotion, or use of literary devices \cite{Heerden2021}. These characteristics point towards the boundaries of current machine creativity, as well as a great potential for expanding their possibilities.

Beyond technical challenges, creative application of AI provokes conceptual inquiries on what constitute creativity, emotional response, and authorship \cite{Wilson1983, Boden1996, Manovich2019}. While their discussion are beyond the scope of this paper, we believe that HITL approaches to develop machine creativity will significantly contribute to such possibilities. Instead of defining or programming creativity, the AI models can learn from our creative outputs which fundamentally contain our emotion, culture, and background. Advanced HITL approaches are meant to increase interaction between human and AI, and this will help generate more diverse, multimodal, and autonomous systems.

\begin{figure*}[tbh!]
\begin{center}
\includegraphics[width=1\textwidth]{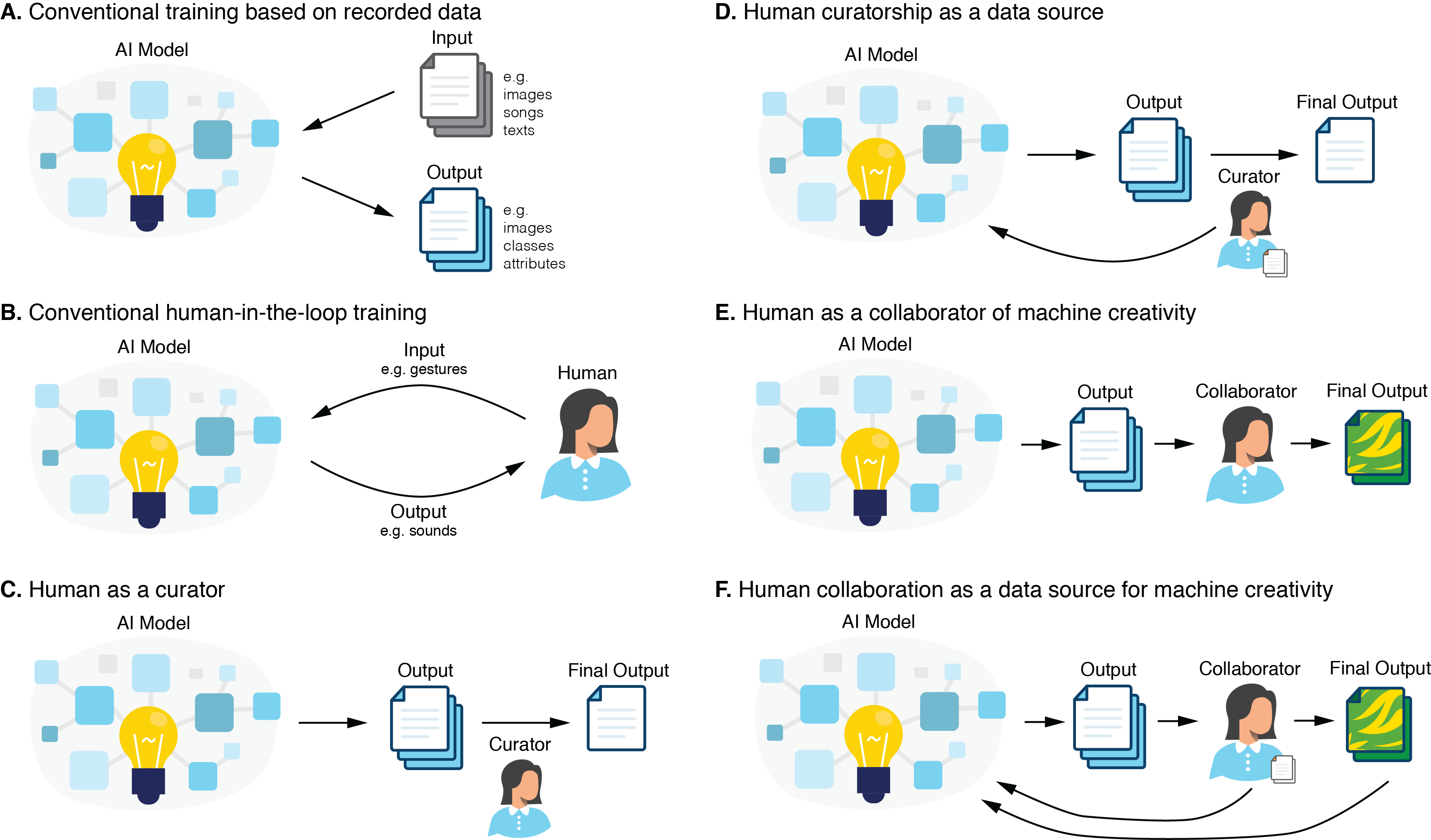}
\end{center} 
\caption{Human in the loop (HITL) to explore and develop machine creativity with AI. \emph{(A)} Conventional training (e.g., neural networks) involves iterating through a large number of recorded data. \emph{(B)} Presently, the most popular use of HITL enables human to give examples of inputs and/or outputs in real-time. \emph{(C)} Human acts as a curator, such that a selection out of a myriad of AI outputs may be exhibited. \emph{(D)} Building on \emph{C}, the curatorship process is fed back into training and generating the output. \emph{(E)} As a collaborator, human interprets and processes the AI outputs to create the final output. \emph{(F)} Building on \emph{E}, we feed what human collaborators are doing and creating back into the AI model. *Inputs, parameters, and other details are omitted.}
\label{hitl}
\end{figure*}
% To generate new outputs (e.g., AI-generated visuals), a model may require inputs such as hyper-parameters, images, texts, and others.

%\section{Existing HITL approaches in creative applications}
%\label{existing}
\section{Conventional Human in the Loop (HITL)}

Presently, there is a great emphasis on end-to-end algorithmic art, and human-AI collaboration is understudied. A current prevailing paradigm in using machine learning, and particularly deep learning, is to train a model by using large data such as paintings of a particular style or theme (Fig. \ref{hitl}A). Such a trained model can reasonably generate new outputs. There has been a growing interest in HITL approaches, where our interactions and feedbacks with the model play a crucial role in training and/or generation \cite{HITLML}. Instead of using a fixed training dataset (Fig. \ref{hitl}A), this conventional HITL approach for creative use places humans in the middle to provide inputs and to guide outputs (Fig. \ref{hitl}B). This can translate our real-time actions (e.g., gestures or brush strokes) into outputs (e.g., musical notes or drawing) \cite{Fiebrink2009}. This presents more flexible and interactive use of AI (Fig. \ref{hitl}B), such immediate feedbacks can create more expressive and usable systems.

\section{Human as Curator}

Another popular -- yet less acknowledged -- use of HITL is to curate outputs of AI for presentation and exhibition (Fig. \ref{hitl}C). As to our best knowledge, all of recent AI paintings involve careful curation by the technologist-artists themselves, even though the importance of this curative process is often minimized or omitted altogether \cite[e.g.][]{GANEdmondBelamy, Elgammal2017}. In fact, in AI art, the distinction among a technologist, an artist, and a curator is blurred, to an extent that \emph{human-as-curator} is mostly assumed. Printing and contextualizing one painting out of seemingly infinite outputs of a trained GAN is likely as important as quality and fidelity of the model. Similarly, writing prompts for a language or image generation models can be also seen as curating both prompts and generated outputs. 
%\section{Proposed HITL approaches}
%\label{proposal}

A computational approach to \emph{human-as-curator} may lead to encoding curatory processes in the system (Fig. \ref{hitl}D). In a simple implementation, human curators mark those selected AI outputs and that information can be used as attributes in further training or priming the network. Conditional GAN \cite{Mirza2014} can be utilized to guide generation, or intermediate layers and their internal representations may help identify features or styles of those curated outputs. A more complex system can be built by incorporating an additional discriminator, which is essentially a connected neural network asking if a given input has been chosen by the human curator. Interestingly, human curators may not be aware or able to explain why they have chosen certain outputs, such that the model may learn unconscious biases and emotions related to those curated set. We foresee that this will lead to new associations linking our biases, emotions, and imaginations, to visual cues, musical notes, and texts.

AI art is not simply an output of an algorithm in its raw form, but its totality which conveys -- or rather create in minds of the audience -- meaning and emotion. Human curators will often work with materials (e.g., what to print on) and environments (e.g., interior design), in which they put the selected AI output. Such artistic and social contexts may become valuable source materials for creative AI models. Connecting our curatory process back to the AI model serves as a first step towards teaching greater contexts that we take for granted. This can be supplemented by providing art theory and history when reasons and themes underlying curatorship are explicit. A full potential of this approach would be realized with greater autonomy in the AI system, including relevant sensors and learning algorithm (e.g., reinforcement learning)

\section{Human as Collaborator}

Instead of a curator, a human could be part of a creation process as a collaborator\footnote{Roles of curators and collaborators might be overlapping, along a spectrum of engagement in art making.}. Specifically, Fig. \ref{hitl}E illustrates humans to base their work on, perform, and process the model outputs. We are witnessing \emph{human-as-collaborator} applications in a wide range of art, ranging from a painting to a film. Collaborative drawing and painting workflows utilize AI and robotic arms to co-create with humans \cite{Jansen2021}. In AIBO, the opera performer's spoken words were used to generate texts from GPT-2 \cite{Pearlman2021}. During a live performance, Time Waves, human artists synthesized and sampled audio and visual elements according to directions from GPT-3 \cite{TimeWaves}. Although not real-time like aforementioned examples, Sunspring \cite{Sunspring} is a short film based on a script written by LSTM \cite{Hochreiter1997}. This LSTM-generated script provided the setting and dialog, which were performed by the director and actors. This process can be seen as general modal translation \cite{Specia2016}, such as from a text to a multimedia or from a musical note to a sound wave. This type of HITL for creative techniques will increase in use, as humans fill in certain roles that AI models are not yet suitable for. In a long term, this flips the conventional roles of humans and computers in which humans dictate and computers support.

The next level of \emph{human-as-collaborator} approach would include humans' interpretative and performative works in further training (Fig. \ref{hitl}F). In aforementioned examples, human artists have played a significant role in creating a final output based on what the computing system has produced. Viewing humans as processing units, we are interested in better understanding and encoding how humans integrate certain inputs and generate artworks. We, as humans, use a variety of personal and cultural contexts -- even subconsciously. While this information is difficult to quantify or summarize, HITL provides a framework to work with our innate creative and emotional responses. For example, with multi-modal architectures such as DALL-E \cite{Ramesh2021} and CLIP \cite{Radford2021} which are trained on a large corpus of images and texts, a human can paint according to a AI-generated text, and that human painting is fed back into the system. We may create more nuanced optimization problems, such as balancing of HITL paintings and initial training set of paintings.

These types of interactions require new user interfaces, network architectures, training schemes, and more. Innovative technical developments will aid in how we create AI art, what it means to encode creativity, and how we conceptualize machine creativity in general. For example, GPT-3 is not open-sourced and only accessed via texts; advanced computer vision requires powerful GPUs for real-time applications. A robust speech interface for GPT-3 based on speech recognition and synthesis would not only result in interesting experiments but also change our perspective on artificial conversations in the long run. Modularization and acceleration of computer vision within ubiquitous computing devices will enable real-time interaction based on analysis of humans (e.g., facial expressions), human outputs (e.g., paintings), and environments (e.g., of a gallery space). 

In future HITL for machine creativity, human emotions and environmental vibes would be approximated, represented, and fed back into the AI model. This will aid in teaching AI how to better process and represent our creative and emotional responses in generative and expressive manners. The AI models would learn more varied, nuanced, and multi-modal understanding; e.g., texts like `sadness' can not only link to `blue' or `minor chords', but also unexpected sonic textures based on more advanced concepts of harmony and composition. The proposed HITL that relies on more fluid and real-time inputs and outputs would push AI models to discover subconscious relationships and create novel artworks. Furthermore, this process will help us rethink what it means for machines to be creative and for our emotional responses to be encoded.

\section{Conclusion}
\label{conclusion}

Creative expressions derived from AI systems, particularly using neural networks, are flourishing. Expensive sales, prestigious exhibitions, persistent coverage by the media, and longterm interest in academia shows that the rise of AI is a pivotal moment in the history of art. A majority of recent AI art relies on computer vision, audio processing, language models, and other algorithms that are excellent at mimicking, combining, and compositing styles learned from training. When human interactions are needed, they are often limited to initial training or curatorship afterwards. Future HITL for machine creativity will extend this interaction, in order to link different modalities, to interpret high-level concepts, and to mirror emotional responses. Humans can work as curators and collaborators to provide our creative and emotional feedback into creating engaging pieces.

For machine creativity to be appreciated and valued, we need to think about how such outputs can relate to aesthetics and invoke emotions. Even in contemporary art, where aesthetics and emotional responses may be unintentionally or intentionally neglected, it is the temporal and societal contexts that make such pieces interesting. Such contexts would include, but not limited to, histories, traditions, heritages, and technological developments, as well as subconscious and individualistic connections. Teaching an algorithm to take into account the vastness of human experience and diversity of culture is likely impossible. Thus, we propose how expressive and creative training using HITL may help teach AI to understand and mimic certain complex associations and responses we take for granted.

Our proposal for present and future HITL for machine creativity attempts to better reflect this reality. Particularly, human technologists and artists are deeply coupled with final outputs. Imparting authorship or responsibility to AI systems has been dubious and challenging. It would be more fitting to describe, for example, how a human acted as a highly important curator that has chosen one digital image out of thousands of GAN outputs. And, in that curatorship and collaboration, humans have embedded incredible amounts of contextual information that may be the missing piece for true autonomous machine creators. This perspective would help foster deeper exploration and development of machine creativity. Through such human-AI interactions, one may envision how AI may learn multi-modal translation, and a general understanding of our emotional responses.

While we may initially focus on creative applications of AI, these proposed approaches can be applicable in wider domains. Expertises of financial advisors or radiologists may not be easy to encode, yet highly important for profitable selection of investments or prognosis for malignant tumors. Nonetheless, how and when to incorporate a narrow and personal set of domain knowledge requires both technical and ethical consideration. Perhaps, it is allowed for an investment bank to develop their own AI model. We may need to think more critically about whether a single cancer center should create (e.g., fine-tune) their own AI model for cancer diagnosis based on their own practices and procedures for making diagnoses from medical images, blood tests, and patient narratives. In this sense, creative applications readily allow experimentation and development of advanced HITL.

Discussion on `creativity' has a long history in cognitive science, computer science, and art \cite[e.g.,][]{Cohen1979, Wilson1983, Boden1996, Nake1998}. As intelligence can neither be easily defined nor be singularly optimized, creativity is a broad and far-flung goal which can't be quantified, even for humans. What is clear is that state of the art AI models have moved far beyond procedural algorithms and drawing machines in terms of novelty and agency. We envision that the proposed HITL approaches hold a great potential to keep moving the needle towards that elusive machine creativity.

\bibliography{hitl_creativity}
\end{document}